\documentclass[10pt,conference,twocolumn]{IEEEtran}
\IEEEoverridecommandlockouts

\usepackage[tbtags]{amsmath} % defines many math commands and
\usepackage{amssymb}  % get, among others, blackboard bold fonts
\usepackage{enumerate}
\usepackage{amsfonts}
\usepackage{subfig}
\usepackage{color}
\usepackage{cite}

\usepackage{url}
\usepackage{hyperref}

\usepackage{epsfig, graphicx, psfrag}

\newtheorem{thm}{Theorem}

\newtheorem{example}{Example}

\providecommand{\thmref}[1]{Theorem~\ref{#1}}

\providecommand{\secref}[1]{Section~\ref{#1}}

\providecommand{\figref}[1]{Figure~\ref{#1}}

\newcommand{\trace}{\mathrm{trace}}

\newcommand{\Comment}[1]{}
\newcommand{\old}[1]{}
\newcommand{\rem}[1]{}

\newcommand{\bX}{{\bf X}}
\newcommand{\bx}{{\bf x}}
\newcommand{\bY}{{\bf Y}}
\newcommand{\bZ}{{\bf Z}}

\providecommand{\comment}[1]{}

\newcommand{\BC}{{\text{BC}}}
\newcommand{\MAC}{{\text{MAC}}}

\newcommand{\AF}{{\text{AF}}}
\newcommand{\DF}{{\text{DF}}}

\newcommand{\CS}{{\text{CS}}}
\newcommand{\PNC}{{\text{PNC}}}

\newcommand{\beqn}[1]{\begin{eqnarray}\label{#1}}
\newcommand{\eeqn}{\end{eqnarray}}
\newcommand{\beq}[1]{\begin{equation}\label{#1}}
\newcommand{\eeq}{\end{equation}}

\providecommand{\Ddef}{\triangleq}

\newcommand{\tbZ}{{\tilde{ \bf Z}}}

\begin{document}
%\pagestyle{plain}

% paper title
%%%%%%%%%%%%%%%%%%%%%%%%%%%%%%%%%%%%%%%%%%%%%%%%%%%%%%%%%%%%%%%%%%%%%%%%%%%%%%%%%%%%%%%%%%%%%%%%%%%%%%%%%%%%%%%%%%%%%%%%%
\title{Physical-Layer MIMO Relaying}

\author{
\authorblockN{Anatoly Khina}
\authorblockA{Dept. of EE-Systems,
TAU\\
Tel Aviv, Israel \\
Email: anatolyk@eng.tau.ac.il}
\and
\authorblockN{Yuval Kochman}
\authorblockA{EECS Dept.,
MIT\\
Cambridge, MA 02139, USA \\
Email: yuvalko@mit.edu}
\and
\authorblockN{Uri Erez\authorrefmark{1}}
\authorblockA{Dept. of EE-Systems,
TAU\\
Tel Aviv, Israel \\
Email: uri@eng.tau.ac.il}
\thanks{$^*$ This work was supported in part by the U.S. - Israel Binational Science
Foundation under grant 2008/455.}
}

% make the title area
\maketitle

% % % \renewcommand{\thefootnote}{}
% % % %% Uri Erez support
% % % \footnotetext{$^*$ This work was supported in part by the U.S. - Israel Binational Science
% % % Foundation under grant 2008/455.}
% % % \renewcommand{\thefootnote}{\arabic{footnote}}

%%%%%%%%%%%%%%%%%%%%%%%%%%%%%%%%%%%%%%%%%%%%%%%%%%%%%%%%%%%%%%%%%%%%%%%%%%%%%%%%%%%%%%%%%%%%%%%%%%%%%%%%%%%%%%%%%%%%%%%%%

\begin{abstract}
The physical-layer network coding (PNC) approach provides improved performance in many scenarios over ``traditional'' relaying techniques or network coding.
This work addresses the generalization of PNC to wireless scenarios where network nodes have multiple antennas.
We use a recent matrix decomposition, which allows, by linear pre- and post-processing,
to simultaneously transform both channel matrices to triangular forms, where the diagonal entries,
corresponding to both channels, are equal. This decomposition, in conjunction with precoding,
allows to convert any two-input multiple-access channel (MAC) into parallel MACs, over which single-antenna PNC may be used.
The technique is demonstrated using the two-way relay channel with multiple antennas.
For this case it is shown that, in the high signal-to-noise regime,
the scheme approaches the cut-set bound, thus establishing the asymptotic network capacity.
\end{abstract}

\section{Introduction}

The capacity region of networks is a long-standing problem in Information Theory. Wireless networks are of special practical interest. Traditionally, such networks were treated at two different levels: A physical-layer local code which translates the wireless channels into ``bit-pipes'', and a network code over the bit network. This separation is in general sub-optimal, as was demonstrated in recent years by \emph{physical-layer network coding} (PNC) approaches.

In the most basic variant of PNC (for single-antenna nodes), relay nodes simply forward their inputs using power adjustment only (``amplify-and-forward'').  This analog-PNC approach helps in opening network bottlenecks, by allowing a node to assist in transmission even if it cannot decode the message, and indeed it is optimal in some cases, e.g., some limits of the parallel relay network \cite{ScheinGallager00,GastparVetterli05}.
However, analog-PNC suffers from noise accumulation: Without decoding, the relays also forward noise. Structured-PNC is an alternative approach which solves both the bottleneck and noise accumulation effects at the same time. It uses structured/linear codes, building on the property that an integer linear combination of codewords is a codeword as well. A relay node may be able to decode the combination codeword, even if it cannot decode the individual messages.
It was first presented by Wilson et al.~\cite{WilsonRelays} for the simple two-way relay channel.
Nam et al.~\cite{Sae-YoungTwo-Way} have shown that such a scheme achieves a rate, within half a bit of the cut-set bound, for any channel coefficients. A more general network is treated by Nazer and Gastpar's ``compute-and-forward'' strategy \cite{ComputeForward}.

Wireless communication widely uses multiple-input-multiple-output (MIMO) techniques to obtain degrees of freedom. It is therefore natural to ask, what can the combination of PNC and MIMO achieve over a network where terminals have multiple antennas. While some works considered MIMO versions of analog-PNC (see, e.g., \cite{BeamFormingAnalogue,AnalogTwoWay2007,AnalogTwoWay2010}), little was done in the context of structured-PNC. In \cite{MIMO-Compute}, Zhan et al. demonstrate that in the two-way relay channel, structured-PNC can gain from multiple relay antennas - but the work is restricted to single-antenna terminal nodes, not allowing to fully enjoy MIMO gains.

In this work we introduce a way to combine structured-PNC and MIMO transmission, where a relay node receives the linear combination of two
transmit nodes via MIMO channels.
The approach we propose builds on a recently introduced matrix decomposition \cite{STUD:SP}, that allows to simultaneously transform two channel matrices into \emph{triangular} form with equal diagonals, by applying unitary operations at both sides.
The transformation is applied to the channel matrices in conjunction with dirty-paper precoding at the transmitters, which yields parallel channels.

An essential property of the matrix decomposition, which allows to independently apply structured-PNC to each subchannel without loss, is that the ratio between the gains of both transmitters is fixed for all the parallel channels.
We note that the two-way MIMO relay channel problem was recently also considered in \cite{KAIST_TwoWay_Allerton10}, where it was proposed to use the the generalized singular value decomposition (GSVD) \cite{VanLoan76} to transform the matrices into triangular form.
However the GSVD fails to achieve a fixed ratio. Thus, in general it yields sub-optimal results.

The rest of the paper is organized as follows. In \secref{sec:problem} we present the problem and the main result. In \secref{sec:scheme} we provide a constructive proof using the aforementioned decomposition. We conclude in \secref{sec:conclusion} by presenting performance comparisons and discussing some extensions.

\section{The Two-Way MIMO Relay Channel}
\label{sec:problem}

The two-way relay channel consists of two terminals and a relay. We define the channel model as follows. Transmission takes place in two phases, each one, w.l.o.g., lasting $N$ channel uses. At each time instance in the first stage, terminal $i$ ($i=1,2$) transmits a signal $X_{i,n}$ and the relay receives $Y_n$ according to some memoryless multiple-access channel (MAC) $W_\MAC(Y|X_1,X_2)$. At each time instance in the second phase, the relay transmits a signal $X_n$ and the terminals receive $Y_{i,n}$ according to some memoryless broadcast (BC) channel $W_\BC(Y_1,Y_2|X)$. Before transmission begins, terminal $i$ ($i=1,2$) possesses an independent message of rate $R_i$, unknown to the other nodes; at the end of the two transmission phases, each terminal should be able to decode, with arbitrarily low error probability, the message of the other terminal. The closure of all achievable pairs $(R_1,R_2)$ is the capacity region of the network.

\begin{figure}[t]
    \centering
    \subfloat[MIMO MAC phase.]
    {
    \label{fig:MAC_phase}
        \psfrag{&T1}{Terminal 1}
        \psfrag{&T2}{Terminal 2}
        \psfrag{&R}{Relay}
        \psfrag{&H1}{$H_1$}
        \psfrag{&H2}{$H_2$}
%        \centering
        \epsfig{file = 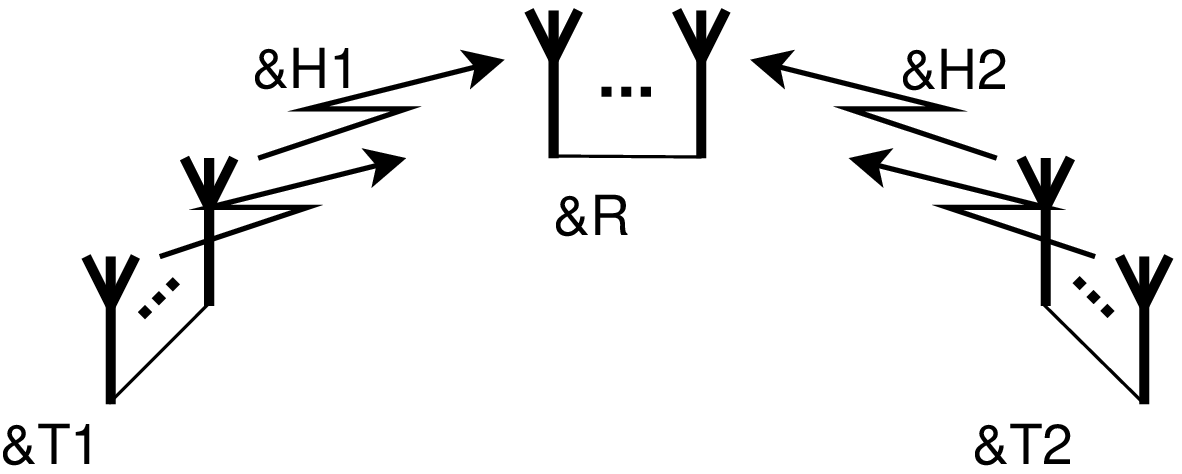, scale = .6}
    }
    \\
    \subfloat[BC phase for the special case where it is a MIMO BC channel.]
    {
    \label{fig:BC_phase}
        \psfrag{&T1}{Terminal 1}
        \psfrag{&T2}{Terminal 2}
        \psfrag{&R}{Relay}
        \psfrag{&H1}{$G_1$}
        \psfrag{&H2}{$G_2$}
%        \centering
        \epsfig{file = 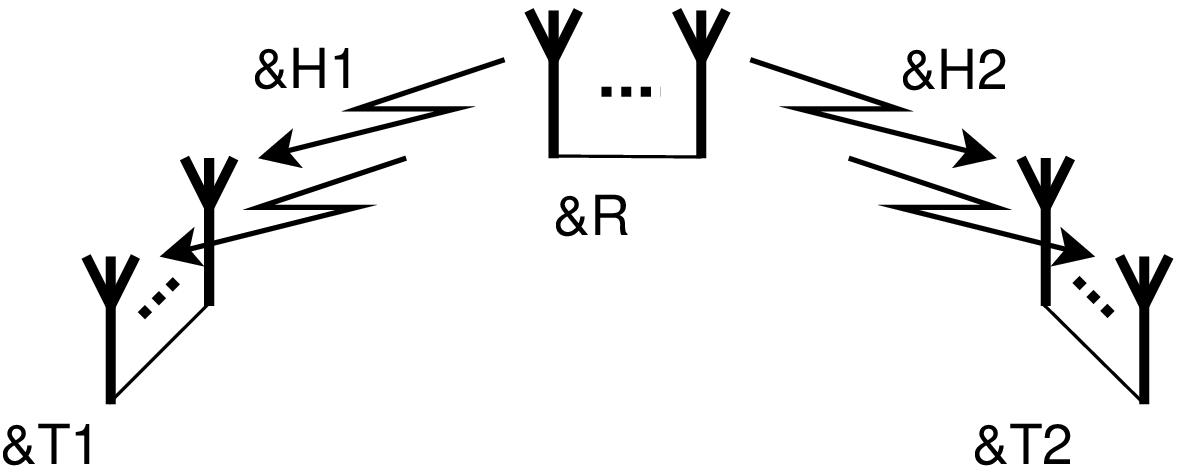, scale = .6}
    }
    \caption{MIMO two-way relay channel. The second (BC) phase may be different in general.}
\label{fig:channel}
\vspace{-3mm}
\end{figure}

We consider a Gaussian MIMO setting, where terminal $i$ ($i=1,2$) has $N_{t;i}$ transmit antennas and  the relay has $N_r$ receive antennas, during the MAC phase, which is depicted in \figref{fig:MAC_phase}. Denoting by bold letters the transmit and receive vectors, we have the MAC channel:
\begin{align*}
\label{MAC}
    \bY & = H_1 \bX_1 + H_2 \bX_2 + \bZ \,,
\end{align*}
where $H_i$ are $N_r \times N_{t;i}$ matrices, $\bZ$ is circularly-symmetric white Gaussian noise with unit variance
and the inputs are subject to the same total power constraint:
\[
    E \left[ \bX_i^\dagger \bX_i \right] \leq P \,, \qquad i=1,2 \,.
\]
We assume that the number of transmit antennas at each node $N_{t;i}$ is at least as large as the number of receive antennas $N_r$, and that the matrices $H_1$ and $H_2$ are full-rank, i.e., have rank $N_r$.\footnote{For the cases in which the matrices are not full-rank or have more receive antennas see \secref{sec:conclusion}.}
We further assume that the messages have the same rate \mbox{$R_1 = R_2 = R$} and that the products of the non-zero singular values, of each of the channel matrices, are equal to $1$, or equivalently:
\begin{align}
\label{eq:det}
    \det \left( H_i H_i^\dagger \right) = 1 \,, \quad i=1,2 \,.
\end{align}

The exact nature of the BC channel is not material in the context of this work. We characterize it using its common message capacity $C_\textrm{common}$.

First consider the single-antenna case, $N_r=N_{t;1}=N_{t;2}=1$. It is shown in \cite{WilsonRelays} that one can achieve a rate of:~\footnote{Wilson et al.~\cite{WilsonRelays} considered a real-valued network, thus an additional $1/2$ pre-log factor was present in their original expression.}
\begin{align}
\label{eq:Wilson_rate}
    R_\PNC = \min \left\{ \log \left(\frac{1}{2} + P \right), C_\textrm{common} \right\} \,.
\end{align}
By the min-cut theorem, one cannot achieve a rate greater than the point-to-point capacity of the MAC links or the common-message capacity of the BC channel:
\beq{eq:SISO-CS} R_\CS = \min \left\{ \log \left(1 + P \right), C_\textrm{common} \right\} \,.
 \eeq It follows that PNC is optimal in the high-SNR limit. It is interesting to compare these to common relaying approaches.
 Using decode-and-forward (D\&F) relaying, the relay must decode both messages with sum-rate $2R$. Instead of forwarding both message it can use a network-coding approach and XOR them, then each terminal can XOR out its own message to obtain the desired one. The resulting rate is given by:
\beq{eq:SISO-DF}
    R_\DF = \min \left\{  \frac{1}{2} \log  \left(1 + 2P \right) , C_\textrm{common} \right\} \,.
\eeq
In the high-SNR limit, when the MAC stage is the bottleneck, the D\&F rate is about half the cut-set bound. In this scenario, an approach that we denote ``amplify-and-forward'' (A\&F) can be used, where the relay forwards the noisy sum of the terminal transmissions.\footnote{Note that it does not necessarily involve analog amplification, if the BC links are not AWGN channels. Instead in that case we compress the sum of the codewords, and treat the unbiased quantization noise as channel noise.}  This approach achieves:
\beq{SISO-AF} R_\AF = \log ( 1 + \alpha P ) \, , \eeq where $\alpha=\alpha(P,C_\textrm{common})$ is a noise amplification factor which is strictly smaller than one for channels of finite capacity. It follows that the A\&F rate has a finite difference from the cut-set bound in the high-SNR limit.

Moving back to the MIMO case, the cut-set bound is given by:

\beq{eq:MIMO-CS} R_\CS = \min \left\{ C_1, C_2, C_\textrm{common} \right\} \, , \eeq where \[ C_i \Ddef \max_{\trace(C_\bx)\leq P} \log \det (I +  H_i C_\bx H_i^\dagger ), \, i=1,2 \,. \]
In the high-SNR limit, under the normalization \eqref{eq:det}, we have: \beq{eq:MIMO-HSNR} \lim_{P\rightarrow\infty} C_i - N_r \log \frac{P}{N_r} = 0. \eeq
The next theorem, which is proved in \secref{sec:scheme}, shows that the PNC rate \eqref{eq:Wilson_rate} generalizes to the MIMO case as follows.

\vspace{.5\baselineskip}
\begin{thm}
\label{thm:two-way}
    The capacity of the Gaussian MIMO two-way relay network with full-rank channel matrices, with $N_t^{(1)},N_t^{(2)} \geq N_r$ satisfying \eqref{eq:det}, is lower-bounded by
    \begin{align}
    \label{eq:Rpnc}
	R_\PNC = \min \left\{ N_r \log \frac{P}{N_r} , C_\textrm{common} \right\} \,.
    \end{align}
\end{thm}
\vspace{.5\baselineskip}

On account of \eqref{eq:MIMO-HSNR}, the rate $R_\PNC$  approaches the cut-set bound for high
SNR as in the single-antenna case. High-SNR conditions hold when
\begin{align}
\label{eq:NonSingularValues_HighSNR}
    \frac{ \left( \lambda_{i;j} \right)^2 P}{N_r} \gg 1 \,, \quad  \begin{array}{l}
                                                                         i=1,2 \\
                                                                         j=1,...,N_r
                                                                       \end{array} \,,
\end{align}
where $\left\{ \lambda_{i;j} \right\}_{j=1}^{N_r}$ are the singular values of $H_i$.

\thmref{thm:two-way} assumes the use of white input. At any finite SNR, it can be improved by optimizing over the input covariance matrices. however, this rate is already better than the rate achievable by the D\&F approach, except for very low SNR.

\section{Structured MIMO-PNC Scheme}
\label{sec:scheme}

In this section we provide a constructive proof for the main
result, stated in \thmref{thm:two-way}. The key ingredient is using the matrix decomposition of
\cite[Sec. IV]{STUD:SP} to obtain equivalent parallel
single-antenna networks, over which we can use the structured-PNC
strategy of \cite{WilsonRelays}. The decomposition is given in
the following theorem of \cite{STUD:SP}, transposed in order to accommodate
for the MAC setting.

\vspace{.5\baselineskip}
\begin{thm}
\label{thm:JET}
    Let $A_1$ and $A_2$ be complex-valued full-rank matrices, of dimensions $m \times n_1$ and $m \times n_2$, respectively,  such that $n_1,n_2\geq m$ (meaning $A_i$ are of rank $m$). If the products of their singular values are equal, then $A_1$ and $A_2$ can be jointly decomposed as
    \begin{align}
    \nonumber
       A_1 &= U T_1 V_1^\dagger \\
    \label{eq:JET_decomposition}
       A_2 &= U T_2 V_2^\dagger \,,
    \end{align}
% % %     where
% % %     \begin{align*}
% % %         a_i \triangleq \sqrt[n]{\prod_{j=1}^n \lambda_{i,j}} \,,
% % %     \end{align*}
    where $V_1$, $V_2$ and $U$ are unitary matrices of dimensions \mbox{$n_1 \times n_1$}, \mbox{$n_2 \times n_2$} and $m \times m$, respectively; and $T_1$ and $T_2$ are generalized lower-triangular matrices (matrices with zero entries above the main diagonal, i.e.,  $T_{1;ij}=0$ and  $T_{2;ij}=0$ for $i<j$, where $T_{i;k,j}$ denotes the $(k,j)$ entry of $T_i$) with positive \emph{equal} diagonal elements.
% % %     and where $\left\{ \lambda_{i,j} \right\}_{j=1}^n$ are the singular values of $A_i$.
\end{thm}
\vspace{.5\baselineskip}

We apply the decomposition of \thmref{thm:JET} to the channel matrices $H_1$ and $H_2$.
% % % \begin{align}
% % % \nonumber
% % %   A_1 &\triangleq \left(
% % % 		       \begin{array}{ll}
% % % 			  H \sqrt{C_{\bX_1}} & I = U T_1 V_1^\dagger
% % %                        \end{array}
% % %                        \right)
% % %   \\
% % %   A_2 &\triangleq \left(
% % % 		       \begin{array}{ll}
% % % 			  H \sqrt{C_{\bX_2}} & I = U T_2 V_2^\dagger
% % %                        \end{array}
% % %                        \right) \,,
% % % \label{eq:AugmentedMatrices}
% % % \end{align}
and denote the diagonal entries of $T_1$ (which are equal to those of $T_2$ by assumption \eqref{eq:det}) by
% $\left\{ t_k \right\}_{k=1}^{N_r}$.
$t_1,\ldots,t_{N_r}$.

We use dirty-paper precoding at each of the terminals to cancel the off-diagonal elements of $T_1$ and $T_2$. This results in $N_r$
parallel channels, with gains given by the (equal for both terminals) diagonal elements. Denote these elements of $T_1$
(which are also equal to those of $T_2$) by $t_1,\ldots,t_{N_r}$. If
we use each such channel as part of an independent single-antenna
two-way relay network with input power $P/N_r$, we can
achieve a total rate of:

\begin{align}
\nonumber
    R_1 &= R_2 = \sum_{k=1}^{N_r} r_k \stackrel{(a)}\geq \sum_{k=1}^{N_r} \log \left(\frac{1}{2} + \frac{t_k^2 P}{N_r} \right) \\
\label{R_k_C_0}
    &\geq  N_r \log \frac{P}{N_r} + \sum_{k=1}^{N_r} \log t_k^2
    = N_r \log \frac{P}{N_r}  \,,
\end{align}
where $r_k$ is the rate conveyed over sub-channel $k$, and $(a)$ holds true due to \eqref{eq:Wilson_rate}.

We now describe in detail a scheme which allows to achieve this rate.

\emph{Codebooks generation}: Apply the decomposition of \thmref{thm:JET} (where we set $n_i=N_t^{(i)}$ and $m=N_r$) to the channel matrices $H_1$ and $H_2$ to obtain matrices $U$, $V_1$, $V_2$, $T_1$ and $T_2$. At each terminal, divide the message into $N_r$ sub-messages of rate
$r_k = R_\PNC/{N_r} + \log t_k^2$.\footnote{If this quantity is negative for some $k$, set $r_k=0$; since this only improves the achievable rate, we can
assume for the sake of analysis that it is always positive.}  Now for each $k$ ($k=1,\ldots, N_r$), both terminals use the \emph{same} nested-lattice
code \cite{ZamirShamaiEreznested} with nesting ratio $r_k$. Denote
the coarse lattice (common for all sub-messages) by $\Lambda$, and
the fine lattices by $\{\Lambda_k\}$. The second moment of
$\Lambda$ is $P/{N_r}$. $\Lambda$ is assumed to be Rogers-good, while
$\{\Lambda_k\}$ are good for AWGN coding.

\emph{MAC encoding}: Terminal $i$ chooses a lattice point $\ell_{i,k}$ for the $k$-th submessage,\footnote{$\ell_{i,k}$ and the signals that follow are $n$-dimensional vectors; we still reserve boldface to denote spatial (antenna) dimension $1,\ldots,N_r$.} then sequentially computes:
\beq{precoder}
    \tilde X _{i,k} = \Bigl[ \ell_{i,k} - \frac{1}{t_k} \sum_{j=1}^{k-1} T_{i;k,j} \tilde X_{i,j} \Bigr] \bmod \Lambda \,.
\eeq
%%where $T_{i;k,j}$ denotes the $(k,j)$ entry of $T_i$.
Finally, the transmit signal is \[ \bX_i = V_i \tilde \bX_i \ . \]

\emph{MAC decoding}: The relay computes $\tilde \bY = U^\dagger \bY$ and then uses lattice decoding to find $\hat \ell_k$, the closest point of $\Lambda_k$ (modulo $\Lambda$), and recovers messages $\{m_k\}$ according to the codebook mapping.

\emph{BC stage}: The relay conveys the messages $\{m_k\}$ to both terminals, using any capacity-approaching common-message BC scheme.
%We note that
%one may use simultaneous equi-diagonal triangularization for this phase as well, using the scheme of \cite{JET:MIMO-BC_Allerton10}.

\emph{Final decoding}: The terminals recover $\{\ell_k\}$, then compute for each submessage:
\beq{unmod}
    \hat \ell_{\bar i, k} = [\hat \ell_k - \ell_{i,k}] \bmod \Lambda \,,
\eeq
where
\begin{align*}
    \bar i \triangleq \left\{ \begin{array}{cc}
                                2 & i=1 \\
                                1 & i=2
                              \end{array}
     \right.
     .
\end{align*}
This estimated codeword is mapped back to a message.

\begin{proof}[Proof of \thmref{thm:two-way}]
    Since $\Lambda$ is Rogers good with second moment $P/N_r$ and $V_i$
    are orthogonal, the transmission satisfies the power constraint.
    Now, \beqn{temp}
        \tilde \bY &=& U^\dagger (H_1 V_1 \tilde \bX_1 + H_2 V_2 \bX_2 + \bZ) \nonumber \\
        &=& T_1 \tilde \bX_1 + T_2 \tilde \bX_2 + \tbZ \,,
    \nonumber
    \eeqn
    where $\tbZ = U$.
    By the precoder operation \eqref{precoder} we have:
    \[
        \left[ \frac{\tilde Y_k}{ t_k} \right] \bmod \Lambda = \left[ \ell_k + \frac{\tilde Z_k}{t_k} \right] \bmod \Lambda \,,
    \]
    where $\ell_k = \left[\ell_{1,k} + \ell_{2,k}\right]\bmod \Lambda$, and $\tbZ$ has the same distribution as $\bZ$.
    Since $\ell_k$ is a valid codeword of $\Lambda_k/\Lambda$,
    this codeword effectively passed through a mod-$\Lambda$ channel with SNR $P_k = t_k^2 P / N_r$.
    Moreover, since $\Lambda_k$ is good for AWGN coding at rate $\log P_k$,
    we have that $\hat \ell_k = \ell_k$ with arbitrarily low error probability (as the dimension $n\rightarrow \infty$).
    Thus, the relay may decode the messages $\{m_k\}$, and the total rate of the latter is \[ R-i =  N_r \log \frac{P}{N_r}\,, \qquad i=1,2 \,. \] Of course, if $C_\textrm{common}$ is lower than that, we can achieve the lower rate.
    Now, these messages can be conveyed to the terminals with arbitrarily low error probability using rate $C_\textrm{common}.$
    If there was no error in any of the stages, then \eqref{unmod} yields
    $\hat T_{\bar i,k} =   T_{\bar i,k}$; by our choice of codebooks,
    this corresponds to a transmission rate $R_\PNC$.
\end{proof}

  \begin{figure}[t]
      \hspace{-5mm}
  %    \centering
      \includegraphics[width=\columnwidth]{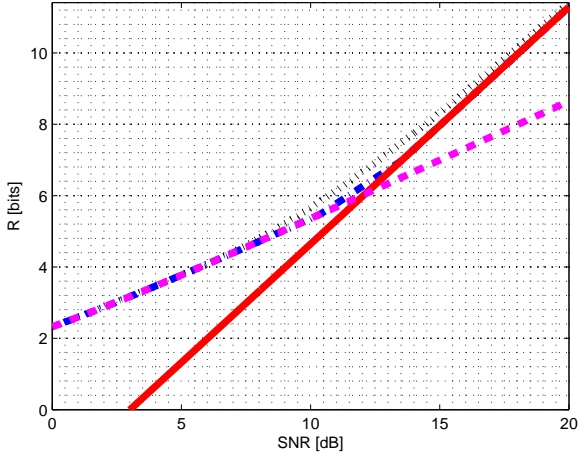}
  %    \vspace{-1cm}
      \caption{Rates as a function of the MAC SNR $P$ in Example~\ref{example} where the MAC section is the bottleneck. Dotted black line~-- $R_\CS$ (cut-set bound); red continuous line~-- $R_\PNC$; dashed magenta line~-- $R_\DF$; dashed-dotted blue line~-- time-sharing between PNC and D\&F.}

      \label{fig:performance}
  \end{figure}

\vspace{3mm}
\begin{example}
\label{example}

  Figure~\ref{fig:performance} depicts the achievable rate obtained when the channel matrices are:
  \begin{align*}
      H_1 = \left(
	      \begin{array}{cc}
		1/2 & 0 \\
		0 & 2 \\
	      \end{array}
	    \right)
	    \,, \quad
      H_2 = \left(
	      \begin{array}{cc}
		2 & 0 \\
		0 & 1/2 \\
	      \end{array}
	    \right)
	    \,,
  \end{align*}
  and the MAC phase is assumed to be the bottleneck, compared to the cut-set bound and to the D\&F rate. As expected, the proposed scheme approach the upper bound as the MAC SNR increases.
\vspace{2mm}

\end{example}

At this point, one can understand why $R_\PNC$ of \thmref{thm:two-way} is higher than that of Yang et al. \cite{KAIST_TwoWay_Allerton10}.  The GSVD-based scheme suggested there induces equivalent triangular channels with \emph{different} diagonals, thus suffering substantial loss when applying the scalar PNC scheme. In fact, as the GSVD provides the ``most spread diagonal ratios'' (see \cite{STUD:SP}), a scheme that does not apply any precoding matrices $U_i$ should have better performance than the GSVD one.

\
\section{Extensions}
\label{sec:conclusion}

\textbf{Time-sharing.}
%As can be seen in \figref{fig:performance},
As is true also for the SISO case, D\&F is better than PNC at low rates;
for example, it yields a positive rate for any positive $P_k$, while PNC fails for $P_k<1$ due to the loss of the ``1" (see
expressions (\ref{eq:Wilson_rate}) and (\ref{R_k_C_0})). As noted for the single-antenna case in \cite{WilsonRelays},
one may use time-sharing in the intermediate region, to improve over both.
\figref{fig:performance} compares the different bounds, as a function of the MAC SNR which is assumed to be the bottleneck in this case.

\vspace{.3\baselineskip}
\textbf{Non full-rank matrices.}
In the case when the channel matrices $H_1$ and $H_2$ are not full-rank or when there are more receive antennas than transmit ones, other strategies needed to accompany the approach proposed in this work. Consider, for instance the following two channel matrices:
%%%%%\begin{align*}
%%%%%    H_1 = \left(
%%%%%            \begin{array}{cc}
%%%%%              a & b \\
%%%%%              0 & 0 \\
%%%%%            \end{array}
%%%%%          \right)
%%%%%          \,,
%%%%%    H_2 = \left(
%%%%%            \begin{array}{cc}
%%%%%              c & d \\
%%%%%              0 & 1 \\
%%%%%            \end{array} \,.
%%%%%          \right)
%%%%%\end{align*}
\begin{align*}
    H_1 = \left(
            \begin{array}{cc}
              1 & 0 \\
              0 & 0 \\
            \end{array}
          \right)
          \,, \quad
    H_2 = \left(
            \begin{array}{cc}
              0 & 0 \\
              0 & 1 \\
            \end{array}
          \right)
          \,.
\end{align*}
In this case the relay sees two parallel channels, from each of the two terminals. Note that in this case a good strategy would be to sum the two channel outputs at the relay, to decode the sum-codeword and send it to the two terminals during the BC phase, or alternately to decode each of the messages, sum them (modulo lattice) and send the result over the BC channel (it is more efficient to send a common message than two private messages). For more general non full rank matrices, such strategies can be combined with the strategy suggested in \secref{sec:scheme} as well applying the joint triangularization of \thmref{thm:JET} to augmtend versions of the channel matrices as is done in \cite{STUD:SP}.

\vspace{.3\baselineskip}
\textbf{Use of non zero-forcing elements.}
We have used the lattice codes
in a ``na\"ive'' manner, achieving $R_k = \log P_k$. That is, we lost the ``entire" $1$.
Indeed, using dithering and MMSE estimation as in \cite{WilsonRelays} one can
achieve the higher rate $R_{PNC,k}$ \eqref{eq:Wilson_rate}. This will
improve performance at low signal to noise ratios (though not
extremely low, as $R_{PNC,k}$ is only positive for $P_k > 1/2$).
Furthermore, instead of the ``zero-forcing'' decomposition proposed in the current work, an MMSE version could be used, which decomposes augmented versions of the channel matrices along with using non-white covariance matrix, as is done in the common message Gaussian MIMO BC case, see \cite{STUD:SP}.
%Indeed, in \cite{JET:MIMO-BC_Allerton10} we show how a broadcast
%scheme based upon the triangularization may be made optimal at any
%SNR. However, applying this scheme is not optimal for rates
%according to \eqref{eq:Wilson_rate}, and it seems that some numerical
%optimization is needed if one is interested in improving the
%finite-SNR PNC rates.

\vspace{.3\baselineskip}
\textbf{Non-symmetric setting.}
The approach can be extended
beyond the symmetric setting, to find achievable rate pairs
$(R_1,R_2)$. Suppose that the ratio between the
singular-values-product of $H_1$ and $H_2$ is $\rho$, then we can
work at each of the parallel networks with ratio $\rho^{1/N_r}$. If
this quantity is not integer, then there is a tradeoff between
rounding and noise amplification \cite{ComputeForward}. However
the loss can be bounded as in \cite{Sae-YoungTwo-Way}.
Furthermore, we may shape the ratios between the diagonals,
such that all the ratios are integers, except for maybe one, or alternately, if the decomposed matrices satisfy a certain majorization condtion,
it is possible to take all the ratios to be equal to 1 except for one, see \cite{STUD:SP} for details.

\vspace{.3\baselineskip}
\textbf{Application: Colored two-way relay channel.}
In the special case where the channel matrices are diagonal, the problem is equivalent to a single-antenna colored channel with piecewise-constant spectrum;
by increasing the matrix dimension, arbitrary spectra can be accommodated for.
Interestingly, the precoding operation at the terminals \eqref{precoder} is in the \emph{frequency domain} in this case,
unlike traditional time-domain Tomlinson precoding.

\vspace{.3\baselineskip}
\textbf{Coding for the broadcast section.}
The only assumption we used regarding the BC section, is that it has a common-message
capacity high enough such that it does not limit performance.
However, it seems likely that the links to the terminal will be
wireless MIMO ones as well. In that case, depicted also in \figref{fig:BC_phase}, the complexity may be
considerably reduced by using the scheme which is based upon the decomposition of \thmref{thm:JET}, for that section
as well; see \cite{STUD:SP}.

%%%%%%%%%%%%%%%%%%%%%%%%%%%%%%%%%%%%%%%%%%%%%%%%%%%%%%%%%%%%%%%%%%%%%%%%%%%%%%%%%%%%%%%%%%%%%%%%%%%%%%%%%%%%%%%%%%%%%%%%%

%\bibliography{mybib}
\bibliography{toly}
\bibliographystyle{unsrt}

%%%%%%%%%%%%%%%%%%%%%%%%%%%%%%%%%%%%%%%%%%%%%%%%%%%%%%%%%%%%%%%%%%%%%%%%%%%%%%%%%%%%%%%%%%%%%%%%%%%%%%%%%%%%%%%%%%%%%%%%%
\end{document}